\def\ifundefined#1{\expandafter\ifx\csname#1\endcsname\relax}

\def\documentstyle{}

\def\title#1\endtitle{\def\thetitle{#1}}
\def\rightheadtext #1\endrightheadtext{\def\therightheadtext{#1}}
\def\author #1\endauthor{\def\theauthor{#1}}
\def\leftheadtext #1\endleftheadtext{\def\theleftheadtext{#1}}
\def\affil#1\endaffil{\def\theaffil{#1}}
\def\address#1\endaddress{\def\theaddress{#1}}
\def\curraddr#1\endcurraddr{\def\thecurraddr{#1}}
\def\email#1\endemail{\def\theemail{#1}}
\def\dedicatory#1\enddedicatory{\def\thededicatory{#1}}
\def\date#1\enddate{\def\thedate{#1}}
\def\thanks#1\endthanks{\def\thethanks{#1}}
\def\keywords#1\endkeywords{\def\thekeywords{#1}}
\def\subjectclass#1\endsubjectclass{\def\thesubjectclass{#1}}
\def\abstract#1\endabstract{\def\theabstract{#1}}

\long\def\references #1\endreferences{%
\bigskip\centerline{References}\bigskip
{\parindent= 0pt \parskip=5 pt #1} }

\def\enddocument{\vfil\supereject\end} 
\def\topmatter{}
\def\endtopmatter{}
\def\thennothing{}


\def \UseCM{
\font\cmrviii=cmr7 
\font\cmrix=cmr9
\font\cmrxii=cmr12
\font\cmrxiv= cmr10 scaled \magstep 2
\def\titlefont{\cmrxii}
\def\smalltitle{\rm}
\def\authorfont{\rm}
\def\headlinefont{\rm}
\def\dedicationfont{\cmrix}
\def\abstractfont{\cmrviii}
\def\footnotefont{\cmrviii}
\rm
}

\newif\iftitle
\def\titlepage{\global\titletrue}
\def\rightrunninghead{\therightheadtext}
\def\leftrunninghead{\theleftheadtext}
\nopagenumbers
\headline=
{\headlinefont\iftitle{}\hfil\else\ifodd\pageno\rightheadline
\else\leftheadline\fi\fi}
\footline=
{\headlinefont\iftitle\hfil\folio\hfil\global\titlefalse\else{}\fi}
\def\rightheadline{\headlinefont\hfil\rightrunninghead\hfil\folio} 
\def\leftheadline{\headlinefont\folio\hfil\leftrunninghead\hfil}

\outer\def\beginsection#1\par
      {\medskip\penalty -100
       \medskip\vskip\parskip
        \message{#1}\leftline{\bf#1}\nobreak\smallskip\noindent
       }

\def\document
{ 
  \parskip=1 pt
  \skip\footins=2\bigskipamount
  \titlepage
  \centerline{\titlefont\thetitle}

\ifundefined{thededicatory}\thennothing\else%
{\smallskip\centerline{\dedicationfont\thededicatory}}
\fi

\bigskip
  \centerline
  {\authorfont\theauthor
    \ifundefined{thethanks}\thennothing\else
     {\footnote{$^1$}{\footnotefont\kern-6pt\thethanks}}
    \fi
  }
\bigskip\medskip
   {\abstractfont
    \baselineskip=9pt
    \leftskip=5pc  \rightskip=5pc
    \noindent
     ABSTRACT.\ \theabstract
\medskip} 
    \ifundefined{thekeywords}thennothing\else
     {\footnote{}{\footnotefont \kern-6pt Keywords:\enspace 
      \thekeywords}}
    \fi
    \ifundefined{thesubjectclass}\thennothing\else
     {\footnote{}{\footnotefont\kern-6pt1980 
     {\footnotefont Mathematics Subject Classification.\enspace}
      \thesubjectclass}}
    \fi
\noindent
}



\UseCM


  \magnification=\magstep1
  \hoffset .65 true in 
  \hsize 26pc 
  \vsize  44pc


\topmatter 

\title  On the Glitch Phenomenon \endtitle

\rightheadtext  On the Glitch Phenomenon \endrightheadtext

\author Leslie Lamport and Richard Palais \endauthor

\leftheadtext L. Lamport and  R. Palais\endleftheadtext

\affil\endaffil 

\address\endaddress

\dedicatory\enddedicatory

\email\endemail 

\date November 8, 1976 \enddate

\thanks 
Research supported in part by NSF Grant No.  NPS 75-08555
and National Software Works contract number F30602-76-C-0094 
\endthanks  

\keywords glitch,synchronization \endkeywords 

\subjectclass \endsubjectclass 

\abstract
 The Principle of the Glitch states that for any device which makes a
discrete decision based upon a continuous range of possible inputs, there
are inputs for which it will take arbitrarily long to reach a decision. The
appropriate mathematical setting for studying this principle is described.
This involves defining the concept of continuity for mappings on sets of
functions. It can then be shown that the glitch principle follows from the
continuous behavior of the device.
\endabstract

\endtopmatter


\def\in{\hbox{\bf I} }
\def\out{\hbox{\bf O} } 
\def\s{\hbox{{\bf S}}} 
\def\I{\hbox{$\cal I$} } 
\def\O{\hbox{$\cal O$} }
\def\S{\hbox{$\cal S$} }
\def\U{\hbox{$\cal U$} }
\def\r{\hbox{\bf R}}     
\def \Definition #1\\ 
{\smallskip
\noindent  {\bf #1. Definition.\quad}}


\document





   There has recently been an increasing awareness of the synchronizer ``glitch''
phenomenon [5,6]. The usual description of this phenomenon states that any
device for deciding which of two asynchronous events occurs first can hang up
in a metastable state for an arbitrarily long time. The purpose of this brief
paper is to indicate an appropriate mathematical setting for describing the
glitch phenomenon and for ``proving'' its existence. The glitch phenomenon 
can be generalized to the following principle:\par

{\narrower\noindent
    For any device making a decision among a finite number of possible 
    outcomes, based upon a continuum of possible inputs, there will be 
    inputs for which the device takes arbitrarily long to reach its decision.\par}

\noindent
It is assumed that the decision making is non-trivial, i.e., that not all possible
inputs to the device lead to the same outcome. A proof of this principle must be 
based upon some continuity assumption about the device, and we will attempt to 
clarify the continuity principles that are involved.\par

   There seem to be three methods by which people attempt to show that glitches
can be avoided: \par

\item{1.} Allowing the device to make an arbitrary decision when it has trouble 
     deciding. This simply introduces additional ``don't care'' decision outcomes,
     and does not help.\par
\item{2.} Introducing noise to drive the device out of its metastable state. The noise
     can be considered to be just an unpredictable input. The introduction of
     noise cannot eliminate the possibility of the device hanging up for an
     arbitrarily long time, but can make it impossible to predict which inputs
     will cause it to do so.\par
\item{3.} Arguing that a decision making device, such as a flip-flop, introduces a 
     discontinuity because it always reaches one of a discrete set of stable final
     states from any of a continous range of initial states. Although the mapping 
     from initial to final states is discontinuous, we will see that this does
     not contradict the basic continuity assumption upon which a proof of the
     glitch phenomenon is based.\par 

\noindent
We now indicate the appropriate formalism for considering the glitch 
phenomenon. Let $\r$ denote the set of real numbers, let $\in$ and $\out$ be two sets,
 and let $\I$ and $\O$ be sets of mappings from $\r$ to $\in$ and from $\r$ to $\out$, 
respectively. An element of $\in$ represents a possible value for the input to the 
device at some instant. An element $i$ of $\I$ represents a possible input to the
device,  where $i(t)$ is the value of the input at time $t$. Similarly, $\out$ is the set of
possible output values, and $\O$ is the set of possible outputs. (For simplicity, we
assume that the device operates for all times. We could also assume that it starts at
some specific time.) The device defines a mapping $\Delta : \I \to \O$; namely  
$\Delta(i)$ is the output produced by the input $i$. In other words, if 
the input at any time $t$ is $i(t)$, then $\Delta(i)(t)$ is the output at time $t$.\par

   As an example, consider an electronic arbitration device with two input wires,
labeled $a$ and $b$, and one output wire, functioning as follows. A single positive voltage
pulse will arrive at each of the two inputs at some times $t_a$ and $t_b$ after time $t=0$ and
before time $t=1$.  If these two input pulses arrive sufficiently far apart, then the device is
to produce a well-defined positive output pulse should the pulse on wire $a$ arrive first, and
the negative of that pulse if the pulse on wire $b$ arrives first. If the two input pulses arrive
closer together than the time resolution of the device, then the device may produce either
the positive or the negative pulse; however, it may  {\it not\/}  produce any type of output
other than the specified positive or negative pulse. Assume for convenience that
all the pulses have the same shape and height, and let $p_s : \r \to \r$ be the continuous
function that represents a positive pulse starting at time $s$, i.e.,  $p_s(t)$ is the 
voltage at time $t$ for such a pulse. To be specific, we assume for the example that 
$p_{s+\epsilon}(t)=p_s(t-\epsilon)$. Then $\in$ is a subset of the  set $\r \times \r$ of
ordered pairs of numbers, and $\I$ is the set of functions of  the form $(p_r,p_s)$ with
$0<r,s<1$,  where $(p_r,p_s)(t)=(p_r(t),p_s(t))$.  The function $(p_r,p_s)$ represents pulses
arriving on wire $a$ at time $r$ and  on wire $b$ at time $s$. The set $\out$ is a subset of
$\r$, and $\O$ is some set of functions from $\r$ into $\r$ containing $\pm p_s$ for some $s$; 
we will see that realistic ``continuous'' behavior of the device implies that $\O$ must also
contain the zero function, which represents the possibility of the device hanging up forever. 
For a more realistic example, we can consider a family of pulses that are acceptable
approximationsto $p_s$. This yields obvious modifications to the sets $\I$ and $\O$. \par 

The above description is purposely made explicit and concrete for purposes of exposition, 
but our argument will apply equally to much more general and abstract devices, provided only 
that they satisfy some basic properties to be made precise below.

   The existence of the glitch is deduced from the {\it continuity\/} of the mapping
$\Delta :\I \to \O$. Continuity is defined for mappings between topological spaces,
and this is the mathematically natural degree of generality in which to derive the results we
shall consider.  However, for simplicity we will restrict our attention to the
special case of {\it metric\/} spaces, and we present next a very brief, self-contained 
account of the few  basic facts about metric spaces that we shall need, referring the reader 
to [2] or any elementary topology text for more details. \par 

   A metric space is a set $X$ of ``points'' in which we have a well-defined 
notion of the distance $d(x, y)$ between any ordered pair of points $(x, y)$ in $X \times X$. 
The distance function $d$ is assumed to satisfy the following three natural properties: 
\item{(i)} $d(x, y) = d(y, x)$;  \par
\item{(ii)} $d(x, y)\ge 0$, with equality only when $x=y$; and \par
\item{(iii)} $d(x, z) \le d(x,y) + d(y, z)$ (the ``triangle inequality'').\par 
\noindent
The most obvious example of a metric space is the set $\r$ of real numbers 
with its ``usual'' distance function $d(x,y)=|x-y|$.
A sequence of points $\{x_n\}$ in $X$ is said to {\it converge\/} to the point $x$ in
$X$ if the sequence $d(x_n, x)$ of real numbers converges to zero, and in this case
we write $x_n \to x$. From (ii) it is easy to see that a given sequence 
$\{x_n\}$ can converge to at most one point $x$, so it makes sense to say that 
$x$ is {\it the\/} limit of the sequence $\{x_n\}$, written 
$x=\lim_{n\to \infty} x_n$. If X and Y are both metric 
spaces and $f: X \to Y$ is a mapping between them, then $f$ is defined to be 
{\it continuous\/} if whenever  $x_n \to x$ in $X$ it follows that 
$f(x_n) \to  f(x)$ in $Y$. Note that this condition can also be written as 
$\lim_{n\to\infty}f(x_n)=f(\lim_{n\to\infty} x_n)$. \par

   Continuity is one of the most important concepts in topology and, as our 
definition shows, it is based on the the notion of convergence.  
We shall assume that our sets $\out$ and $\in$ are metric spaces, so the 
notion of continuity is defined for the mappings $i: \r \to \in$ and 
$\Delta(i) : \r \to \out$. Our sets $\I$ and $\O$ are 
assumed to be sets of {\it continuous\/} functions, and we now want to define what it
means for the mapping $\Delta:\I \to \O$ to be continuous. This of course 
requires defining the appropriate notion of convergence in $\I$ and $\O$. 

Let us assume more generally that we are given a metric space $\s$ with a distance
function  $d$ and also a  set $\S$ of continuous functions from $\r$ into $\s$. 
To simplify notation, given a positive number $r$ let us 
define $d^r(f,g)=\max_{-r\le t\le r}d(f(t),g(t))$ (i.e., the maximum
distance between corresponding points on the graphs of $f$ and $g$ in the interval
$[-r,r]$).  We will say that a sequence $\{f_n\}$ in $\S$ converges to an element $f$ in $\S$.
if for each positive $r$, $f_n$ converge to $f$ {\it uniformly\/} on the interval
$[-r,r]$---this means that  $d^r(f_n(t),f(t))$ should converge to $0$ for each $r$. 
Following standard mathematical terminology, we shall also refer to this mode of 
convergence as ``convergence in the compact-open topology''. 
It is not hard to  check that $d^*(f,g)= \sum_{r=1}^\infty 2^{-r} d^r(f,g)/(1+d^r(f,g))$
defines an  explicit distance function $d^*$ for $\S$ such that $d^*(f_n,f)\to 0$ if and
only if $f_n$ converges to $f$  in the compact-open topology. 

(To get some feeling for this mode of convergence, let us take $\s=\r$, and take for $\S$ all
continuous functions $f:\r \to \r$. Let $f_0$ denote  the identically zero function. Define
$f_n(t)$ to be zero if $t$ is less than $0$ or  greater than $2/n$, $f_n(t)=n t$ for  $0 \le t
\le 1/n$, and  $f_n(t)=2-n t$ for $1/n \le t \le 2/n$. Since $f_n(t)=0$ when  $n>2/t$ it is
clear  that $f_n$ converges to $f_0$ pointwise, but $d^1(f_n,f_0)=1$ so the convergence is not 
uniform on $[-1,1]$. However, if we take $f_n(t)=e^{x-n}$, then since each $f_n$  is
monitonically increasing, it follows that  $d^r(f_n,f)=f_n(r)=e^r e^{-n}$, which tends to zero 
as $n \to \infty$, so $f_n$ {\it does\/} converge to $f_0$ in the compact-open
topology, even though each $f_n(t)$ tends rapidly to infinity with $t$.) \par   

   As an example, let $\s$ be the set $\r$ of real numbers, and let $p_s$ be the
element of $\S$ of our example above. We assume that $p_s(t)$ equals zero except when $t$ is
in some finite interval (depending on s). Then the continuity of the function
$p_s$ together with the relation $p_{s+\epsilon}(t)=p_s(t-\epsilon)$ 
implies the expected result that if $s_n$ converges to $s$ in $\r$, then $p_{s_n} \to p_s$ in
the compact-open topology of $\S$. That is, the mapping $F:\r\to \S$ defined by $F(s) =p_s$ is
continuous.  A more surprising result of our definition is that if $s_n \to \infty$,  then 
$p_{s_n} \to 0$ (where $0$ denotes here the identically zero function).
 This follows from the fact that if $p_s(t)=0$ for all $t<r$,  then $d^r(p_s,O) = 0$.
 
   We will need one more concept from topology before we are ready to discuss the
glitch phenomenon. Let $U$ be a subset of a metric space $S$, and let $[0, 1]$ 
as usual denote the interval of real numbers $t$ with $0 \le t \le 1$.
If $u_0$ and $u_1$ are points in $U$, then a path in $U$ from $u_0$ to $u_1$ is a
continuous mapping $\pi: [0, 1] \to S$ such that $\pi(0) = u_0$, 
$\pi(1) = u_1$, and $\pi(t)$ is in $U$ for all $t$ in $[0,1]$. We
say that $U$ is a {\it pathwise connected\/} subset of $S$ if such a $\pi$ can be found
for each choice of $u_0$ and $u_1$ in $U$. 
If $F$ is a continuous mapping of $U$ into a metric space $T$ and $\pi:[0,1] \to \U$
is as above, then the composition $F\circ \pi:[0,1] \to T$ is a path in $T$ from $F(u_0$) to
$F(u_1)$. It follows that if $U$ is a pathwise connected subset of $T$, then $F(U)$ is pathwise
connected  subset of $T$, where $F(U)$, the image of $U$ under $F$, is the set of all points in
$T$ of the form $F(u)$ for some point $u$ in $U$. \par 

 Let $\S$ again be the space of all continuous functions from $\r$ into $\r$, and let
$\U_r$ be the set of all functions $\pm p_s$ with $0 < s \le r$. If we set 
$p_\infty=0$, this also defines $\U_\infty$. Now one can show that $\U_r$ 
is {\it not\/} pathwise connected for any $r$ with $0 <r <\infty$, e.g., there is no
continuous path $\pi: [0, 1] \to \U_{100}$  such that $\pi(0) = p_1$ and 
$\pi(1)=-p_1$. However, $\U_\infty$ {\it is\/} pathwise connected. For example, we can
define a continuous path $\pi: [0, 1] \to \U_\infty$   with$\pi(0) = p_1$ and 
$\pi(1)=p_1$ as follows:  \par 

$$ \pi(s)=\cases{p_{1/(1-2s)},      & if $0 \le s < 1/2$;\cr
                            0,      & if $s=1/2$;\cr       
                 p_{1/(2s-1)},      & if $1/2 \le s \le 1$.\cr}$$

If we let $\U_r$ be the set of all pulses that are sufficiently close to $\pm p_s$
for some $0<s \le r$, then we again find that $\U_r$ not pathwise connected 
unless $r = \infty$.
(This assumes that no pulse is close to both $p_s$ and $-p_{s'}$, unless $s=s'=\infty$.)

 To establish the Principle of the Glitch, one proves three things:

\item{1)} The mapping $\Delta: \I \to \O$ is continuous, using the compact-open 
          topologies on $\I$ and $\O$. \par

\item{2)} The space $\I$ is pathwise connected. \par 

\item{3)} The set of outputs in $\Delta(\I)$ for which the decision is made 
          before some fixed time r is not pathwise connected.  \par 

\noindent                          
Since 1) and 2) imply that $\Delta(\I)$ is pathwise connected, 3) shows that for any
finite time $r$ there must be inputs in $\I$ for which the device does {\it not\/} 
reach a decision by time $r$. \par

   Parts 2) and 3) of the proof are usually easy. In our example, to prove that $\I$
is pathwise connected, we must construct a continuous path $\pi : [0, 1]\to \I$ with
$\pi(0)=(p_{r_0},p_{s_0})$ and $\pi(0)=(p_{r_1},p_{s_1})$, for any $r_1$ and $r_0$ 
between $0$ and $1$. This can easily be done by taking  
$\pi(t) = (p_{(1-t)\,r_0+t\,r_1}, p_{(1-t)\,s_0+t\,s_1})$.  Part 3 is proved by showing
the $\U_r$ is not pathwise connected for $0<r<\infty$. \par

   To prove that $\Delta$ is continuous with respect to the compact-open topologies, one
must make some assumption about the nature of the dynamical equations that govern the
behavior of the device. If one assumes that $\in$ and $\out$ are finite-dimensional
manifolds, and that the behavior of the device is described by a system of ordinary
differential equations (possibly involving delays) then one need only make fairly
natural assumptions about these equations in order to deduce the continuity 
of $\Delta$. The reader is referred to [1,3,4] for the appropriate theorems.

   The mathematical situation is not so satisfactory if $\in$ and $\out$ are infinite 
dimensional and the device is described by a system of partial differential
equations. We know of no general result that can be applied in this case.

   Any proof of the existence of the glitch phenomenon raises the question of
the extent to which a mathematical result can be applied to a physical situation.
This is a metaphysical question that is beyond the scope of this paper. We
merely observe that science is based upon the assumption that an approximately
correct theory will describe the approximate behavior of a system. If the proof
is based upon a theory that closely approximates the physical device, then we
may safely conclude that the device must occasionally take very much longer
than usual to make a decision. Whether it must ``really'' take an unbounded
length of time to decide cannot be determined from any approximate theory.

\references
[1] E. A. Coddington and N. Levinson. Theory of Ordinary Differential Equations. 
McGraw Hill, New York, 1955.

[2] J. Dieudonn\'e. Foundations of Modern Analysis. Volume 1, Academic Press,
   New York, 1960. 

[3] J. Hale.  Functional Differential Equations.  Springer-Verlag, New York,
   1971.

[4] M. Hirsch and S. Smale. Differential Equations, Dynamical Systems and
   Linear Algebra. Academic Press, New York, 1974.

[5] M. P\v echou\v cek.  Anomalous response times of input synchronizers.  IEEE
   Transactions on Computers, 25:133-139, February 1976.

[6] T. J. Chaney and C. E. Molnor. Anomalous behavior of synchronizer and
arbiter circuits. IEEE Transactions on Computers, 22:421-422, April 1973.

\endreferences

\enddocument